\documentclass[a4paper,11pt]{article}
\usepackage{aaskaiid}
\usepackage{orcidlink}
\title{Ultra-Precise Astrometric Search for Exoplanets with SKA-VLBI}
\ShortTitle{SKA-VLBI Exoplanets}

\author[1]{Salvador Curiel\orcidlink{0000-0003-4576-0436}}
\ShortName{Curiel et al.} 
\author[2,3]{Mar\'ia J. Rioja\orcidlink{0000-0003-4871-9535}}
\author[2]{Richard Dodson\orcidlink{0000-0003-0392-3604}}
\author[4,5]{Huib Jan van Langevelde\orcidlink{0000-0002-0230-5946}}
\author[6]{Tao An\orcidlink{0000-0003-4341-0029}}

\affiliation[1]{Instituto de Astronom{\'\i}a, Universidad Nacional Aut\'onoma de M\'exico (UNAM), Apartado Postal 70-264, Ciudad de M\'exico, M\'exico}
\emailAdd{scuriel@astro.unam.mx}
\affiliation[2]{International Centre for Radio Astronomy Research (ICRAR), University of Western Australia, 35 Stirling Hwy, Crawley, WA 6009, Australia}
\affiliation[3]{Observatorio Astron\'omico Nacional (IGN), Alfonso XII, 3 y 5, 28014 Madrid, Spain}
\affiliation[4]{Joint Institute for VLBI ERIC (JIVE), Oude Hoogeveensedijk 4, 7991 PD Dwingeloo, The Netherlands}
\affiliation[5]{Leiden Observatory, Leiden University, Postbus 2300, 9513 RA Leiden, The Netherlands}
\affiliation[6]{Shanghai Astronomical Observatory, CAS, Nandao Road 80, Shanghai 200030, China}

\abstract{
The study of exoplanets is a rapidly developing field, driven by the discoveries of Kepler and TESS, among others. The recent detection of Jovian planetary companions of low-mass stars demonstrates that VLBI observations will be an excellent tool for indirect detection of planetary companions through precise radio astrometry of the host star. The anticipated sensitivity of SKA-VLBI and its capability to form multi-beam Tied Array Beams and MultiView analysis will allow us to achieve an order of magnitude increase in astrometric precision, providing much finer details for a wider range of exoplanets and hosts, which will revolutionize the field of exoplanets. Precise micro-arcsecond astrometric observations are crucial for detecting not only Jupiter-like planets, but also lower-mass planets. SKA-VLBI astrometric observations in L and C bands will open the possibility of indirect detection of thousands of planetary companions to radio-bright ultra cool dwarfs, M dwarfs and young stars. When a companion is also detected,  the astrometric fit of the data will provide the dynamical masses of the components. In the case of binary systems with planets, fitting the astrometric data will provide the individual masses of stars and  planets, as well as the mutual inclination angle of the system, which will show whether the planet is moving in prograde or retrograde orbit around its host star. The search for exoplanets at radio wavelengths will be complementary to other techniques and will allow for the detection of a population of exoplanets that is difficult to reach using other techniques.
}

\begin{document}
\newcommand{\actaa}{Acta Astron.} 
\newcommand{\araa}{ARA\&A} 
\newcommand{\aar}{A\&ARv} 
\newcommand{\aapr}{A\&ARv} 
\newcommand{\ab}{Astrobiol.} 
\newcommand{\aj}{AJ} 
\newcommand{\apj}{ApJ} 
\newcommand{\apjl}{ApJL} 
\newcommand{\apjs}{ApJSS} 
\newcommand{\ao}{Appl. Opt.} 
\newcommand{\apss}{Astro. \& Space Sci.} 
\newcommand{\aap}{A\&A} 
\newcommand{\aaps}{A\&AS.} 
\newcommand{\baas}{Bull. Am. Astron. Soc.} 
\newcommand{\caa}{Chinese A\&A} 
\newcommand{\cjaa}{Chinese J. A\&A} 
\newcommand{\cqg}{Class. Quantum Gravity} 
\newcommand{\gal}{Galaxies} 
\newcommand{\gca}{Geo. Cosmo. Acta} 
\newcommand{\icarus}{Icarus} 
\newcommand{\jcap}{JCAP} 
\newcommand{\jgr}{J. Geophys. Res.} 
\newcommand{\jgrp}{J. Geophys. Res. Planets} 
\newcommand{\jqsrt}{J. Quant. Spectrosc. Radiat. Transf.} 
\newcommand{\memsai}{Mem. SAIt} 
\newcommand{\mnras}{MNRAS} 
\newcommand{\nat}{Nature} 
\newcommand{\nastro}{Nat. Astron.} 
\newcommand{\ncomms}{Nat. Commun.} 
\newcommand{\nphys}{Nat. Phys.} 
\newcommand{\na}{New Astron.} 
\newcommand{\nar}{New Astron. Rev.} 
\newcommand{\physrep}{Phys. Rep.} 
\newcommand{\pra}{Phys. Rev. A} 
\newcommand{\prb}{Phys. Rev. B} 
\newcommand{\prc}{Phys. Rev. C} 
\newcommand{\prd}{Phys. Rev. D} 
\newcommand{\pre}{Phys. Rev. E} 
\newcommand{\prx}{Phys. Rev. X} 
\newcommand{\prl}{Phys. Rev. Let.} 
\newcommand{\psj}{Planet. Sci. J.} 
\newcommand{\planss}{Planet. Space Sci.} 
\newcommand{\pnas}{Proc. Natl Acad. Sci. USA} 
\newcommand{\procspie}{Proc. SPIE} 
\newcommand{\pasa}{PASA} 
\newcommand{\pasj}{PASJ} 
\newcommand{\pasp}{PASP} 
\newcommand{\rmxaa}{RMXAA} 
\newcommand{\sci}{Science} 
\newcommand{\sciadv}{Sci. Adv.} 
\newcommand{\solphys}{Sol. Phys.} 
\newcommand{\sovast}{Soviet Ast.} 
\newcommand{\ssr}{Space Sci. Rev.} 
\newcommand{\uni}{Universe} 

\maketitle

\section{Introduction} 
In the field of exoplanet research, the study of radio emission from M-dwarfs and ultra-cool dwarfs (UCDs) is of relevance for several reasons. First, magnetic interactions between a star and a close-in planetary companion are expected to produce observable radio bursts as the companion traverses the stellar magnetosphere  \citep[e.g.,][]{Strugarek2025}. The mechanism responsible for these radio bursts is the electron-cyclotron maser instability (ECMI), which produces highly polarized and strongly beamed emission \citep[e.g.,][]{Pena2025}.
Second, high-precision astrometry of radio-bright M-dwarfs and UCDs can be used to search for companions with masses in the planetary-mass regime. For example, the astrometric precision of the Very Long Baseline Array (VLBA; $\lesssim$ 100 $\mu$as in X-band, single epoch) allows the indirect detection of massive planets in wide orbits (P $\gtrsim$ 100 days) around nearby (d $\lesssim$ 10 pc) M-dwarfs and UCDs \citep[][]{Curiel2020,Curiel2022}. Multi-epoch radio monitoring of such systems can then be used to characterize their orbits as well as their masses, providing a means to accurately estimate the physical properties of potential exoplanet-hosting stars. Moreover, high-angular-resolution observations in the radio can detect close-in stellar and even substellar companions around this kind of stars, providing the possibility of obtaining their dynamical masses \citep[e.g.,][]{Curiel2024}. 
Last, high-energy charged particles confined in a dipolar magnetic field around very late M-dwarfs (ultracool dwarfs) can produce radiation belts that show a double-lobed structure \citep[][]{Kao2023,Climent2023}. These structures are similar in morphology to the Jovian radiation belts and emit detectable non-thermal (synchrotron) radio emission, which can be used to probe planet-like magnetospheric phenomena  \citep[][]{Kao2023}.
The sensitivity and resolution of Very Long Baseline Interferometry (VLBI), in combination with the discovery of radio emission from late M and L dwarfs \citep[][]{Berger2001,Berger2006}, have become a game changer to improve our observational understanding of planetary formation in very low-mass stars and brown dwarfs with M $<$0.2 $M_{\odot}$ \citep[see][]{Bower2009,Forbrich2009,Forbrich2013,Gawronski2017}. In addition, recent work has demonstrated the potential of the VLBI astrometric technique to discover brown dwarf and planetary companions. For instance, using astrometric observations at radio wavelengths with the Very Long Baseline Array (VLBA),  three brown dwarfs were discovered orbiting two pre-main sequence stars \citep{Curiel2019, Curiel2024}, and two Jovian mass planets were found associated with an M9 ultracool dwarf and the main star in an M-dwarf binary system \citep{Curiel2020, Curiel2022}.

Some of these topics are discussed in other chapters in Advance Astrophysics with the Square Kilometre Array II, such as magnetic star-planet interaction {\color{black}(Radio Emission from Star-planet interactions by \citealp[][]{Vedantham01.2026.SKA})}, direct exoplanet detection {\color{black}(Discovery and characterizing exoplanets and ultracool dwarfs with the Square Kilometre Array by \citealp[][]{Kavanagh01.2026.SKA}}) and Ultra-Precise MultiView Astrometry {\color{black}(Ultra-Precise Astrometry with the next generation of VLBI, using SKA by \citealp[][]{Rioja01.2026.SKA})}. In this chapter, we provide an overview of the core science goals and requirements for exoplanet astrometric studies, along with a focus on the advances expected from the next generation astrometric MultiView methods to meet this challenge, and recent progress made from pathfinder studies. In particular, we focus on what is expected from SKA in conjunction with other radio telescopes, located at distances of between several hundred and several thousand kilometers, SKA-VLBI.

\section{Micro arcsecond Astrometric precision with SKA-VLBI}\label{sect:Precision_SKAVLBI}

\subsection{Next generation MultiView methods}

VLBI intrinsically provides the highest resolution in astrometry and, with a robust calibration method, provides an astrometric precision roughly equal to the resolution over the signal-to-noise ratio (S/N). Nevertheless, in current VLBI arrays and using standard single-calibrator phase reference methods, precision is nearly always limited by systematics in the calibration rather than intrinsic thermal limits \citep[see][Figures 2 and 3]{RiojaDodson2020}.
The SKA offers the possibility for ultra-precise astrometry
in conjunction with other radio telescopes, to provide next-generation VLBI observations with potential for an order of magnitude improvement in sensitivity and in astrometric performance.
Furthermore, the wide range of radio frequencies that will be covered will allow for the application of astrometric methods to a much wider set of targets and science cases.
But to deliver this, we require, in addition to the next-generation instruments, next-generation
methods, to improve on the limits currently obtained with standard astrometric methods. 
MultiView methods have demonstrated outstanding performance, with an increasing number of astrometric measurements reaching the thermal noise limit of current VLBI networks, as predicted by our error analysis. As discussed in Chapter Ultra-Precise Astrometry \citep[][]{Rioja01.2026.SKA}, we estimate an order of magnitude improvement for SKA-VLBI  (i.e. up to 10$\mu$as astrometric precision at 1.4 GHz and 1$\mu$as above 5 GHz), assuming an upgraded network of antennas that match the SKA capabilities combined with next-generation technology. 
This is predicated on the sensitivity improvement from the increased collecting area and the quasi-perfect compensation of systematic atmospheric effects, as provided by MultiView analysis using multiple calibrators or an extremely close suitable calibrator (with an arcmin separation or less). Additionally, the possible frequency range for astrometry could increase by an order of magnitude. 

\begin{figure}[ht]
    \centering
	\includegraphics[width=0.49\columnwidth]{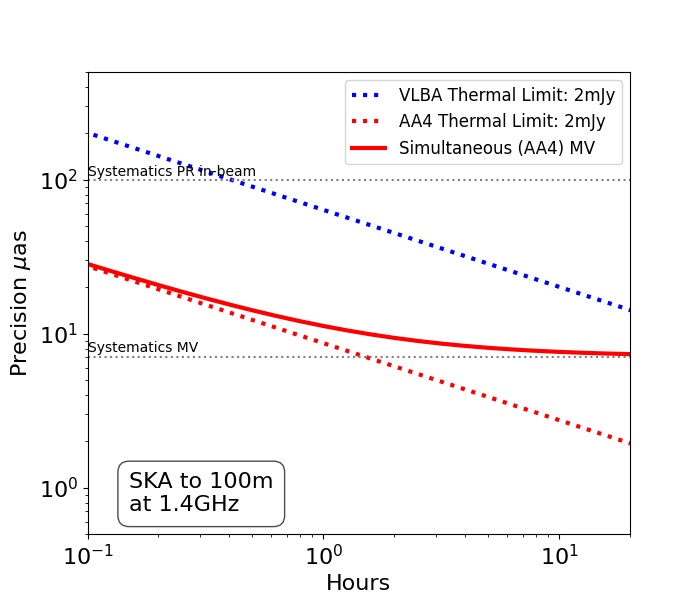}
	\includegraphics[width=0.49\columnwidth]{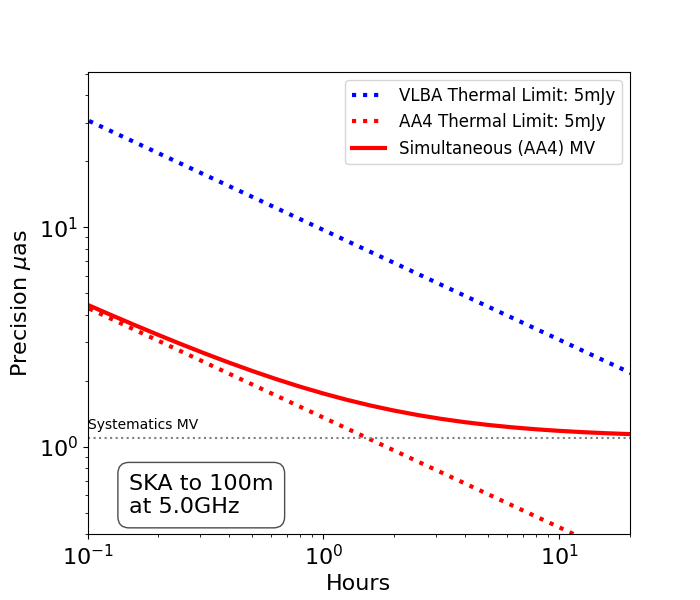}
    \caption{Astrometric Error budget for SKA with multiple pencil beams at AA4 sensitivities. (Left) 1.4 GHz for a 2 mJy source and (Right) 5 GHz for a 5 mJy source. Thermal limits for VLBA (blue dashed) and for an SKA-VLBI single baseline to a nominal 100 m-diameter telescope, are calculated using SEFDs from EVNCalc (red dashes for AA4 tied-array of 10 km radius) as a function of the observing time. At L-band the systematics limits for a single very close phase reference calibrator (6 arcmin away) and MultiView with multiple sources are shown with black dotted lines, at 100 $\mu$as and 7 $\mu$as respectively. For C-band only MV calibration is shown (as few high dynamic range in-beam calibrators are to be found at 5 GHz), limiting at 1 $\mu$as for simultaneous. The predicted performance (solid lines) is a sum of the systematic and thermal errors. At the selected target flux systematic limits are reached at about an hour; weaker or stronger sources would reach the same limits in greater or lesser time, respectively.
    }
    \label{fig:Errors-astrometry_2}
\end{figure}

\subsection{Consequences of SKA Capabilities}
One of the most ground-breaking enhancements the SKA-VLBI will bring is the ability to form multiple Tied Array Beams (TAB) from the SKA array, both within the large field of view of a single element with maximal sensitivity or, with sub-arraying, anywhere within the visible sky.
The initial roll-out of SKA (AA*) will be limited to one (or maybe four) simultaneous beams. 
However, it remains possible that the Pulsar Timing Voltage beams can be repurposed for VLBI, at the price of reduced bandwidth.
The prime driver for multiple VLBI beams is for the ultra-precise cancelation of systematic errors, but it would also allow for targeting of multiple sources close together. 
One can estimate the impact of such an observing model by including dynamic errors in the error budget, see Figure \ref{fig:Errors-astrometry_2}. See also the Chapter Ultra-Precise MultiView Astrometry \citep[][]{Rioja01.2026.SKA}.

\section{Planetary companions detectable with SKA-VLBI radio astrometry}\label{sect:Num_planets}
Section~2 shows that the expected MultiView astrometric precision of SKA-VLBI observations at 1.4 and 5 GHz is between 7 and 1 $\mu$as, respectively. According to the expected performance of SKA1 \citep{Braun2019} the expected sensitivity at those frequencies is about 2 and 1.3 $\mu$Jy/beam in 1 h integration time.
This extraordinary performance will open up the possibility of detecting numerous exoplanets using radio astrometry. To quantify the number of exoplanets that could be detected with SKA-VLBI using this technique, we first need to estimate the number of low-mass stars (radio-emitting or not)  that can be detected with an adequate S/N. We used a 3D Monte-Carlo population synthesis simulation to estimate the number of  expected detections within a volume with a radius of 300 pc around the Sun, which is consistent with estimates of the Galactic thin disk scale height, as reported values range from 150 and 450 pc \citep[]{Aganze2022}. The estimated local density of ultra-cool dwarfs (spectral types M7-T6) is assumed to be 2.9$\times10^{-2}$ pc$^{-3}$, consistent with volume-limited ($<$25 pc) surveys \citep[][and references therein]{Tang2022}  and  the completeness-corrected 20 pc census of \citet{Kirkpatrick2024}. 
Unfortunately, while the stellar space density of M-dwarfs is well established, the local fraction of radio-bright M-dwarfs at the relevant frequencies is not yet uniformly quantified for a robust estimate of the local density of M-dwarfs, so we will limit these calculations to ultra-cool dwarfs.
Since there are no estimates of the expected visible sky, for these calculations, we consider that half of the sky will be accessible to be observed with SKA-VLBI. We assume the detection rate of radio-flaring ultra-cool dwarfs from unbiased surveys, $f_{rad}$ = 3$\%$. Each synthetic source is given a radio luminosity following the luminosity function of ultra-cool flares \citep{Tang2022} and their flux densities are calculated at 1.4 and 5 GHz. After running 10000 simulations, we obtain that $\sim$18000 ultra-cool dwarfs could be detected at 1.4 GHz and $\sim$2700 at 5 GHz as far as 300 pc with a 10$-\sigma$ detection limit.

Until now, only two exoplanets have been found using radio astrometry with the VLBA: a Saturn-like planet associated with the M9 Dwarf TVLM~513-46546 located at a distance of 10.762$\pm$0.027 pc,  and the other one a Jupiter-like planet associated with the main star in the M-dwarf binary system  GJ~896AB located at a distance of 6.257$\pm$0.007 pc \citep{Curiel2020, Curiel2022}. Since very few exoplanets have been detected so far using radio astrometry, we lack an estimate of the detection rate of planets associated with radio-emitting ultra-cool dwarfs. As a first approach, we consider a  rate of UCDs with at least one planetary companion $f_{plan}$ = 3$\%$, consistent with the detection rate of planets from RV surveys.
However, the rate of  planets  associated with radio-emitting ultra-cool dwarfs could be higher than 3$\%$. Under these assumptions, we find that $\sim$550 and $\sim$82 ultra-cool dwarfs with planetary companions as far as 300 pc could be detected with SKA-VLBI observing at 1.4 and 5 GHz, respectively. SKA-VLBI monitoring of these UCDs could lead to astrometric detection of their planetary companions.

\begin{figure}[ht]
    \centering
	\includegraphics[width=0.45\columnwidth]{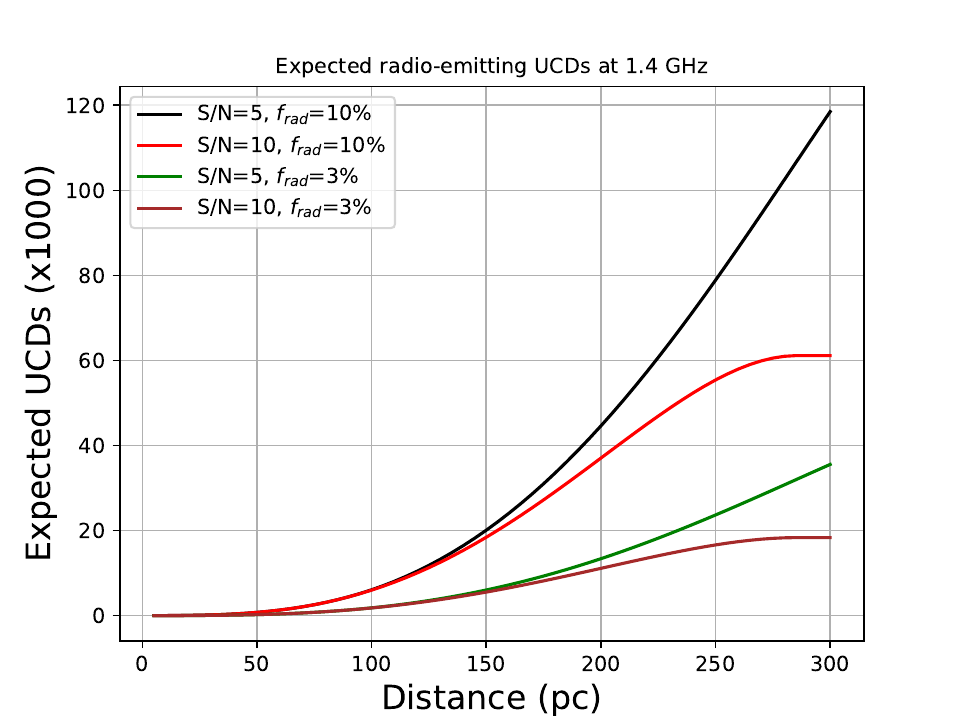}
	\includegraphics[width=0.45\columnwidth]{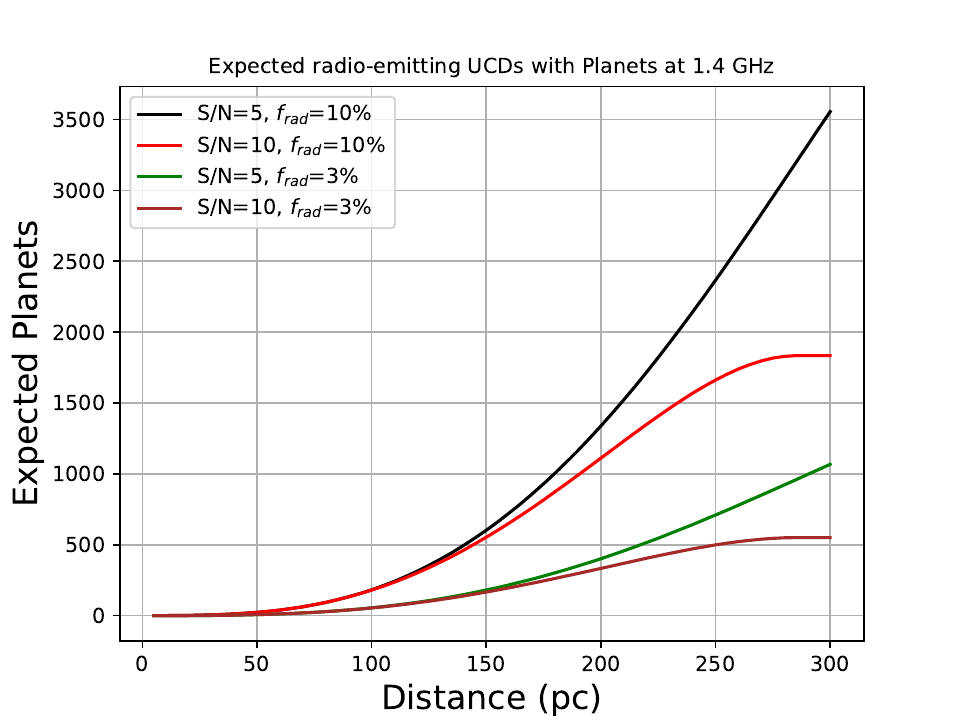}
	\includegraphics[width=0.45\columnwidth]{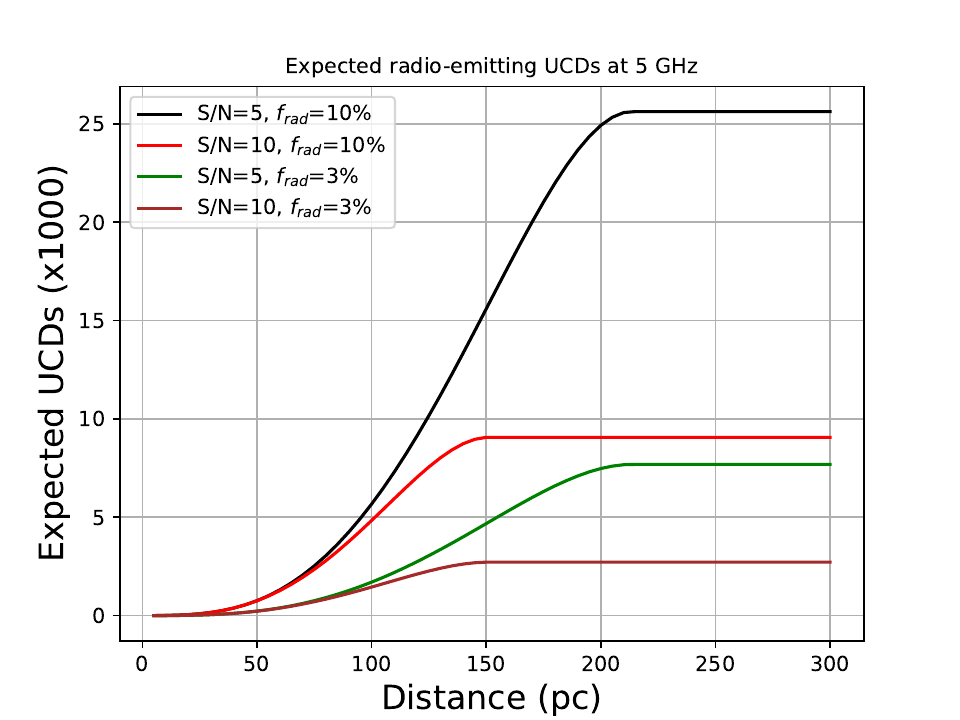}
	\includegraphics[width=0.45\columnwidth]{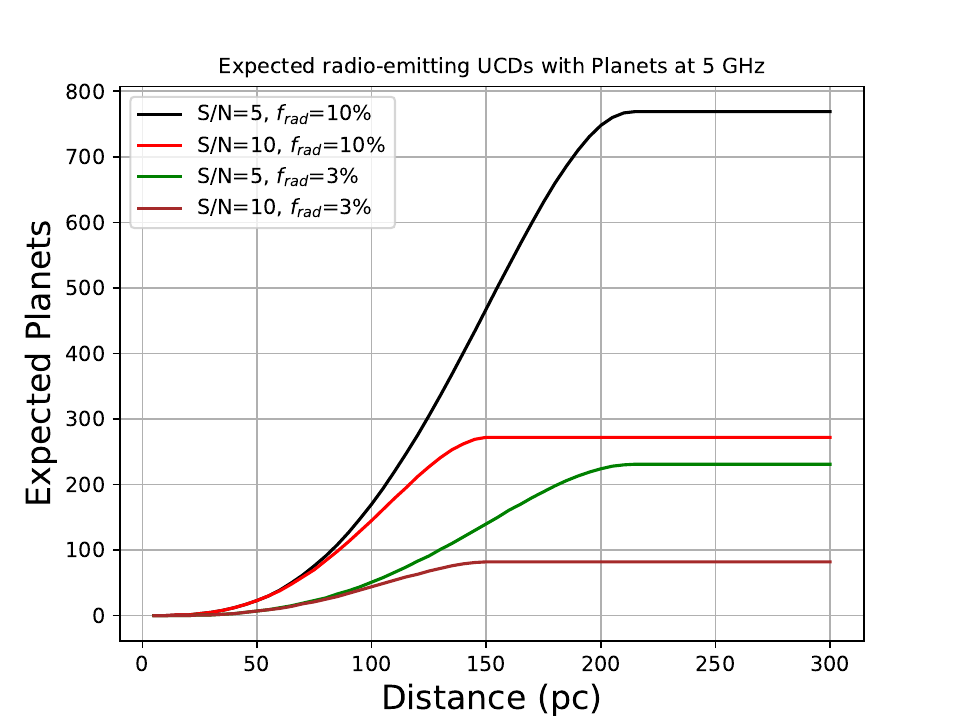}
    \caption{SKA-VLBI astrometry at 1.4 and 5 GHz. Left-panels: Cumulative expected radio-emitting UCDs detectable with SKA-VLBI.
                 Right-panels:  Cumulative expected radio-emitting UCDs with Exoplanets detectable with SKA-VLBI. Given the non-thermal origin of the emission, more UCDs with planetary companions could be detected at low frequencies than high frequencies. In addition, at low frequencies it will be possible to detect UCDs with planetary companions at larger distances.}
    \label{fig:UCDs_distance}
\end{figure}

\begin{figure}[ht]
    \centering
	\includegraphics[width=0.45\columnwidth]{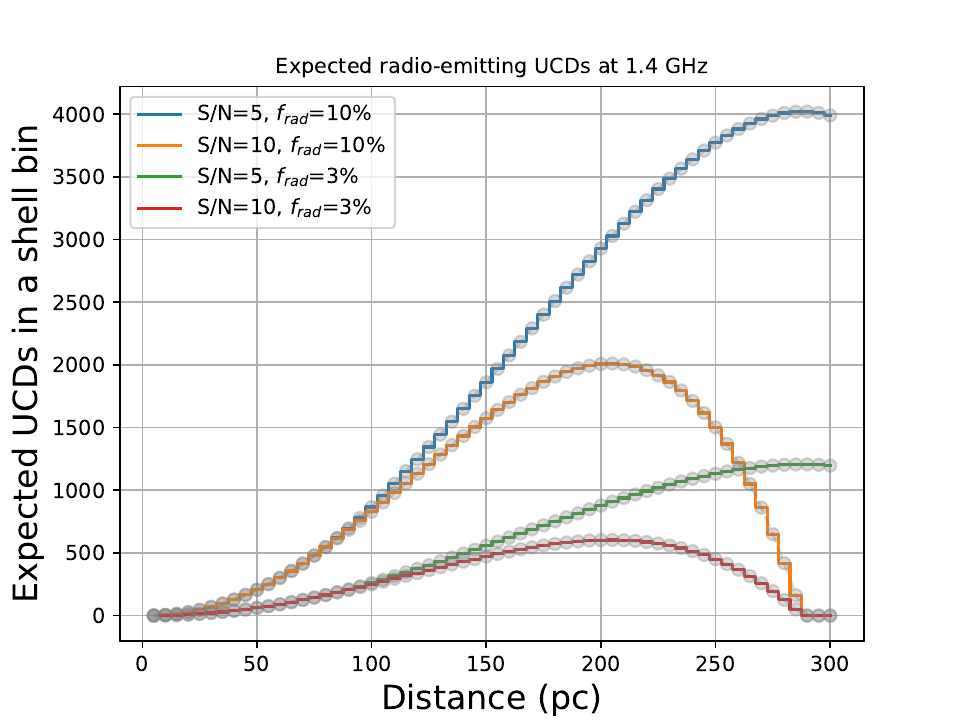}
	\includegraphics[width=0.45\columnwidth]{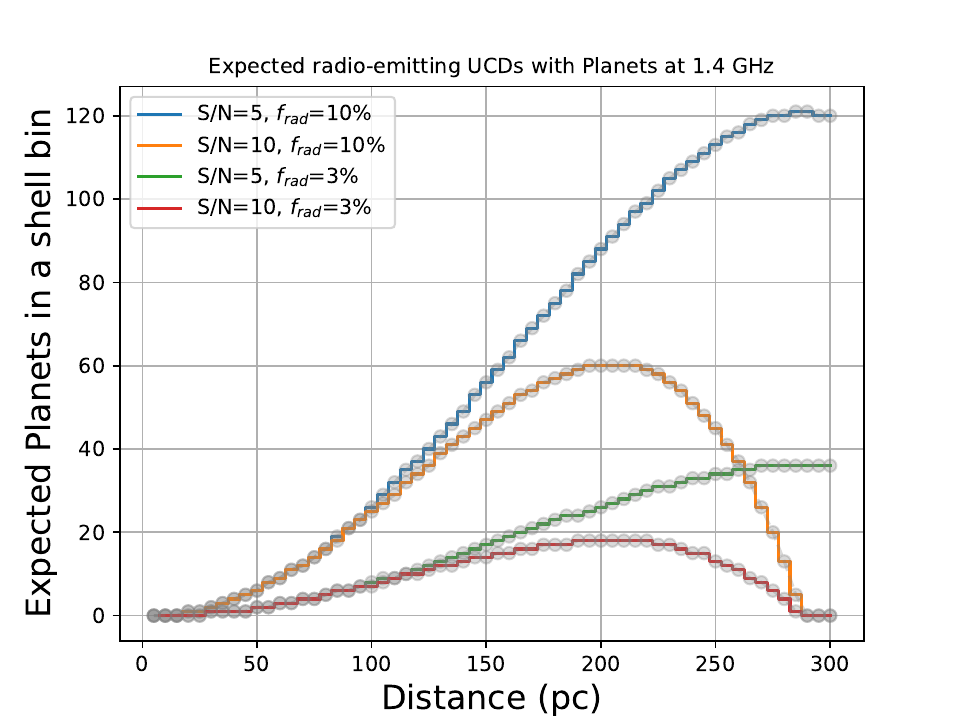}
	\includegraphics[width=0.45\columnwidth]{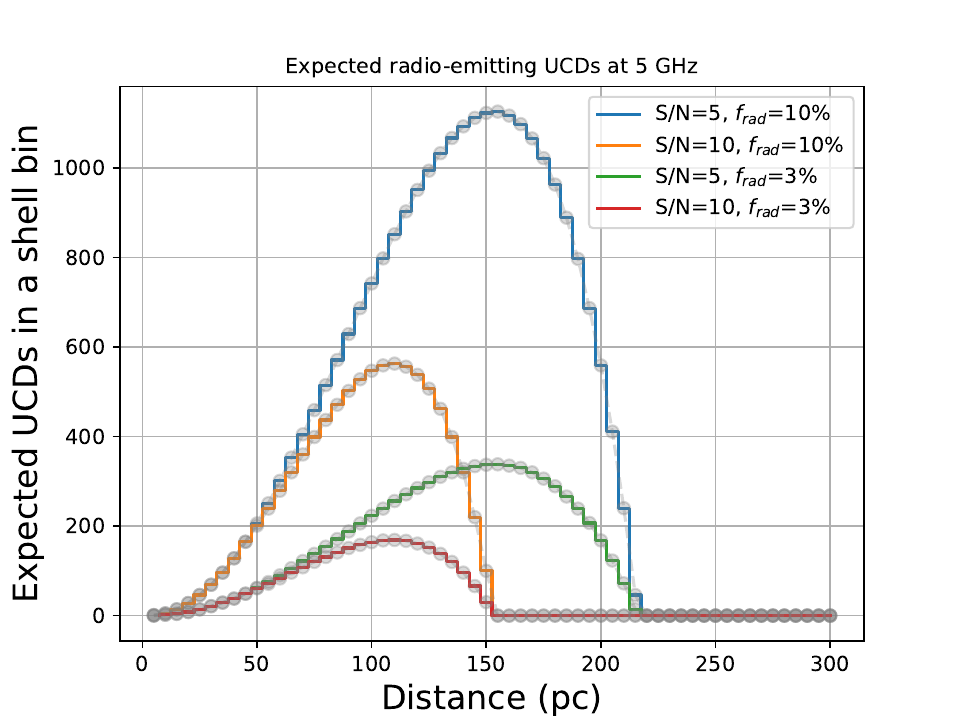}
	\includegraphics[width=0.45\columnwidth]{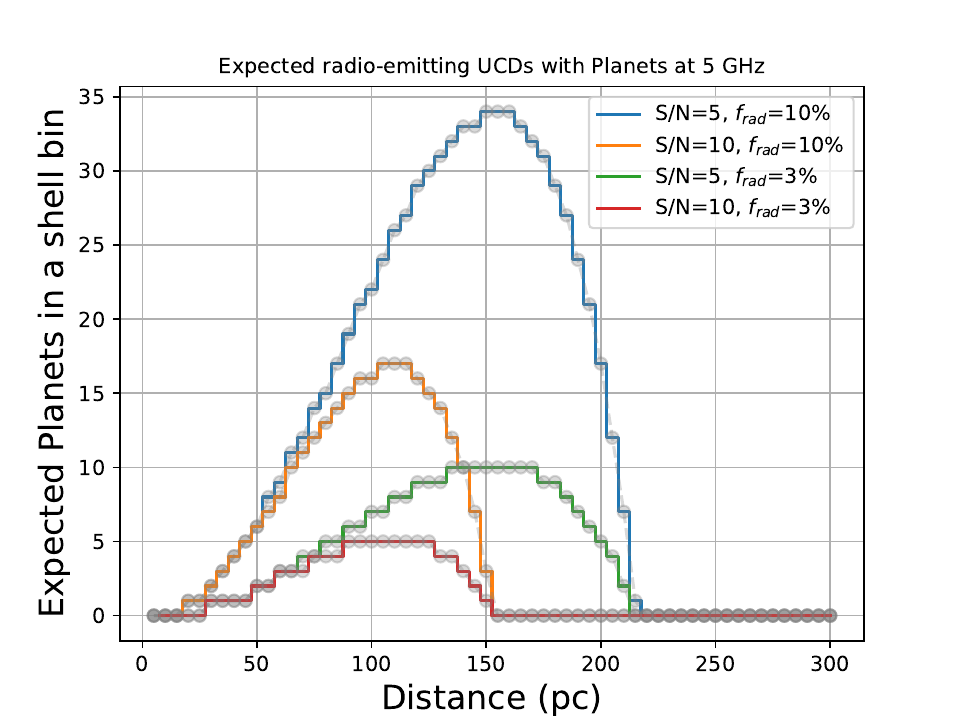}
    \caption{Left-panels:  Expected radio-emitting UCDs in the shell bins detectable with SKA-VLBI.
                 Right-panels:  Expected radio-emitting UCDs with Exoplanets detectable in each shell bin with SKA-VLBI. Given the non-thermal origin of the emission, there is a distance limit after which UCDs with planetary companions cannot be detected. In addition, at low frequencies it will be possible to detect UCDs with planetary companions at larger distances.}
    \label{fig:UCDs_distance_bin}
\end{figure}

Table \ref{tab:Expected_plan} presents the expected number of radio-emitting ultra-cool dwarfs and ultra-cool dwarfs with planetary companions under different scenarios, including a detection rate typically quoted for ultra-cool dwarfs of $f_{rad}$ = 10$\%$.  
A sensitivity of 4 $\mu$Jy/beam was used at both frequencies. 
Figure \ref{fig:UCDs_distance} shows the cumulative number of expected radio-emitting UCDs, as well as  the cumulative number of expected radio-emitting UCDs  with exoplanets at different distances. Figure \ref{fig:UCDs_distance_bin} shows the number of radio-emitting UCDs  with exoplanets in each of the spherical shells used in the 3D Monte-Carlo population synthesis simulation. This figure shows that radio-emitting UCDs  with exoplanets could be found beyond the 300 pc limit used in these calculations, which could significantly increase the number of exoplanets that could be found with SKA-VLBI precise radio astrometry.

The relevant parameters were chosen mainly to yield a conservative estimate of the expected number of radio-emitting UCDs with planets. Here, we do not take into account that earlier M-dwarfs and young stars can also be detected at radio frequencies. A large number of M-dwarf and TTauri stars have already been detected with the VLA and ATCA, and, in some cases, also with the VLBA and EVN. Using multi-epoch VLBA observations, a Jupiter-like planet was found associated with an M3.5 star, and two low-mass Brown Dwarfs (BD) were found associated with a young  weak-line TTauri star, one of them was a direct detection and the other using radio astrometry \citep[e.g.][]{Curiel2019, Curiel2022}. There are no estimates of the space density of these kinds of stars. However, these are also potential targets for the search for exoplanets using radio astrometry with SKA-VLBI. The search for planets associated with this kind of stars will considerably increase the number of planets that could be found with multi-epoch SKA-VLBI observations. 

\begin{table}[ht]
	\centering
	\caption{Expected number of radio-emitting ultra-cool dwarfs, and radio-emitting  ultra-cool dwarfs  with exoplanets that could be detected with SKA-VLBI at 1.4 and 5 GHz and a sensitivity of 4~$\mu$Jy~beam$^{-1}$.}
	\label{tab:Expected_plan}
	\begin{tabular}{rrrrrr} 
		\hline
		Freq (GHz) & S/N & $f_{rad}$ & $f_{plan}$ & $N_{radioUCD}$ & $N_{plan}$\\
		\hline
		1.4 &   5 &   3$\%$ & 3$\%$ & 3.6$\times$10$^{4}$ & 1067 \\
		1.4 & 10 &   3$\%$ & 3$\%$ & 1.8$\times$10$^{4}$ & 550 \\
		1.4 &   5 & 10$\%$ & 3$\%$ & 1.2$\times$10$^{5}$ & 3556 \\
		1.4 & 10 & 10$\%$ & 3$\%$ & 6.1$\times$10$^{4}$ & 1834 \\
		5.0 &   5 &   3$\%$ & 3$\%$ & 7.7$\times$10$^{3}$ & 231 \\
		5.0 & 10 &   3$\%$ & 3$\%$ & 2.7$\times$10$^{3}$ & 82 \\
		5.0 &   5 & 10$\%$ & 3$\%$ & 2.6$\times$10$^{4}$ & 769 \\
		5.0 & 10 & 10$\%$ & 3$\%$ & 9.1$\times$10$^{3}$ & 272
	\end{tabular}
\end{table}
 
 \section{Precise Astrometry as a tool to search for exoplanets}
\begin{figure}[ht]
    \centering
	\includegraphics[width=0.45\columnwidth]{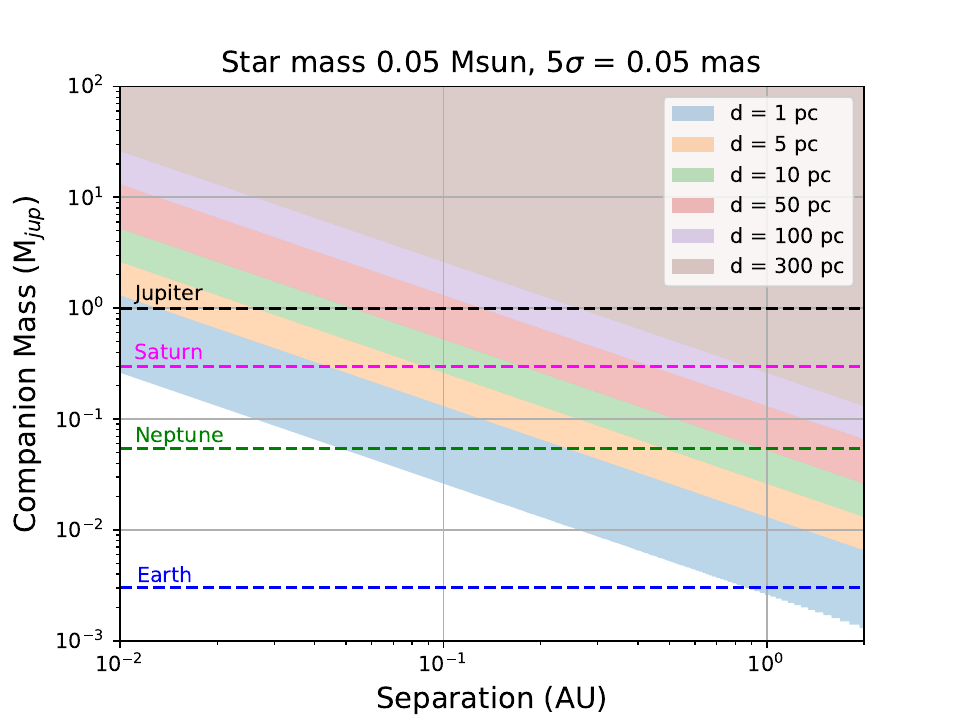}
	\includegraphics[width=0.45\columnwidth]{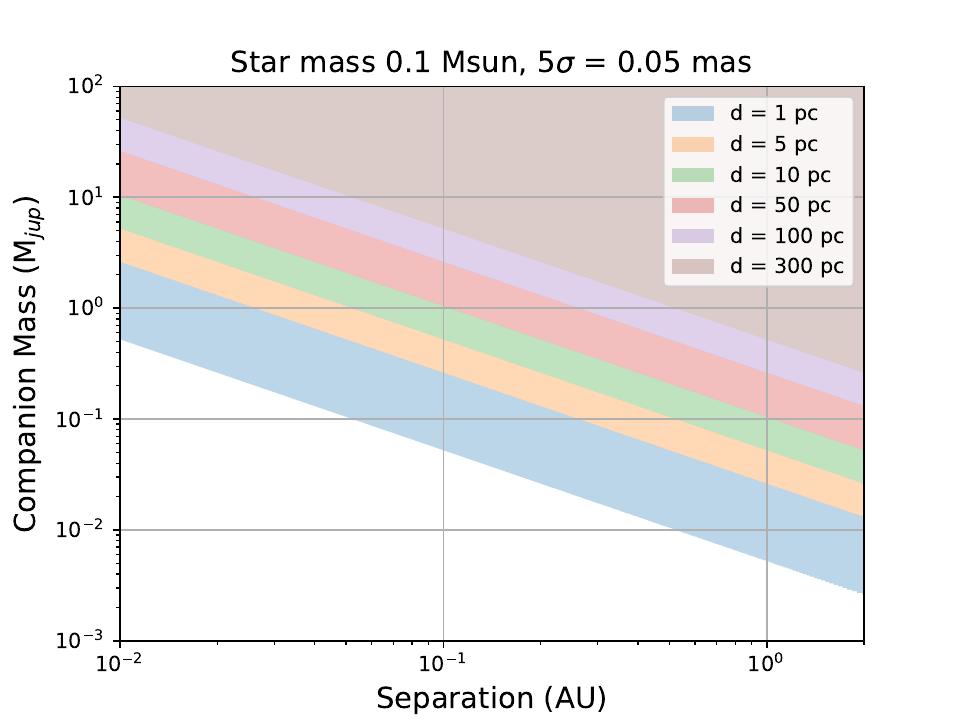}
	\includegraphics[width=0.45\columnwidth]{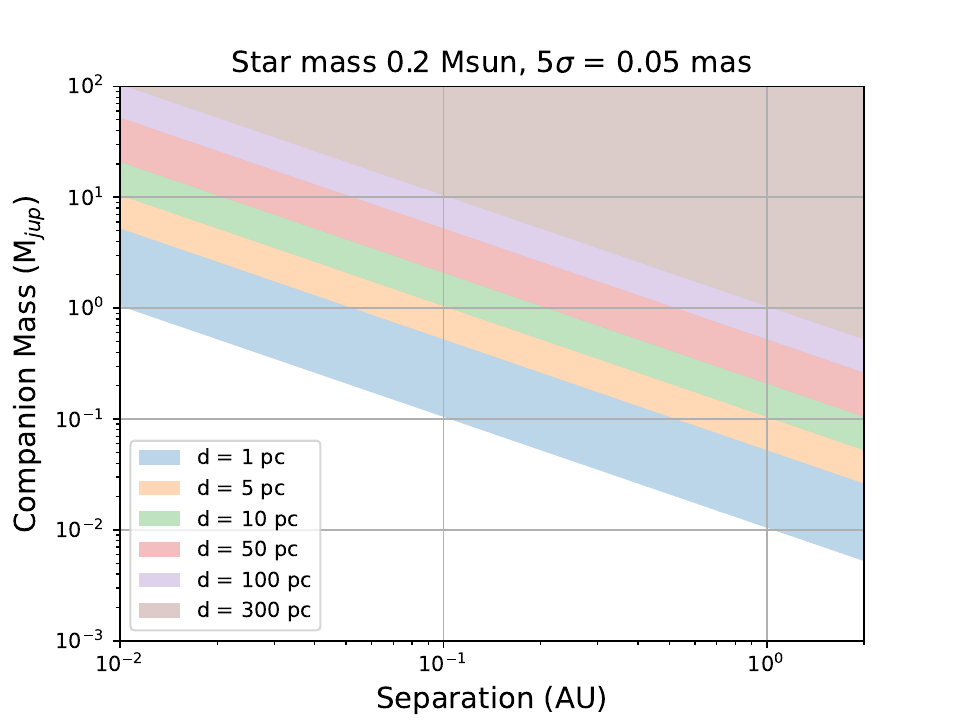}
	\includegraphics[width=0.45\columnwidth]{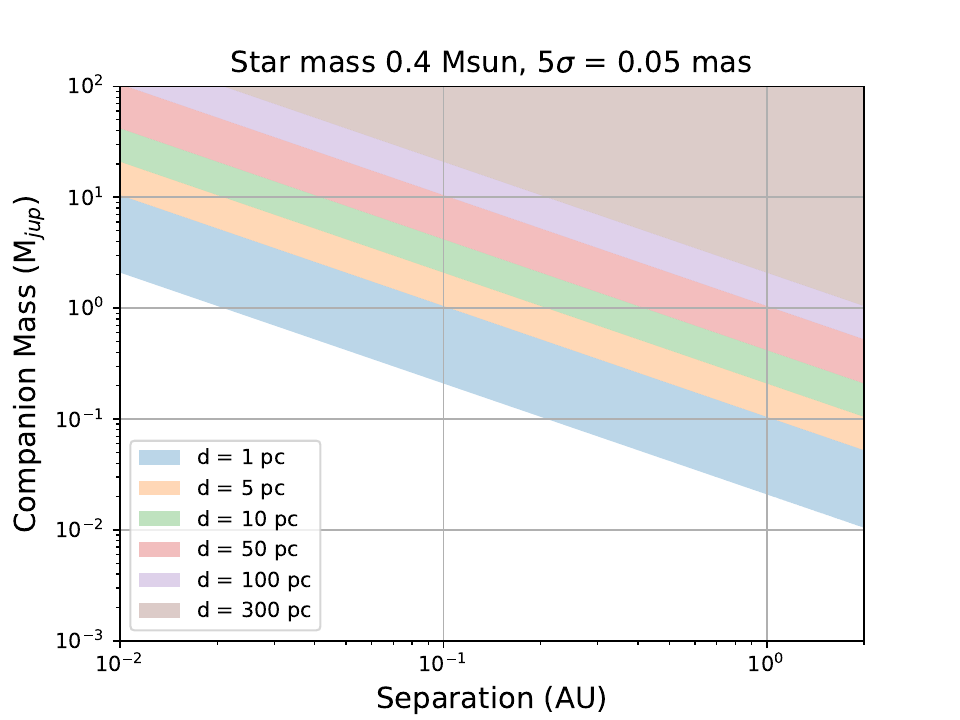}
    \caption{SKA-VLBI astrometry. Expected astrometric signal for a combination of stellar mass and distance, produced by a wide range of planetary masses and semi-major axes (orbital periods). The color bands correspond to the expected minimum astrometric signal of 0.05 mas ($\sim$5$\sigma$) for a distance of 1, 5, 10, 50, 100 and 300 pc. 
 The lower end of the color band marks the 5$\sigma$ (50~$\mu$as) astrometric signal threshold at each distance. For any distance, only companions that produce an astrometric signal above this 5$\sigma$ limit would be detectable with the SKA-VLBI. 
As a reference, the horizontal color lines correspond to the Solar System planets. 
This figure shows that at a given distance, lower-mass planets will be detected around lower-mass stars, and that Jovian planets will be detected around stars with masses up to 0.4 M$\odot$, even at distances above 100 pc.
The SKA-VLBI will be able to detect Jovian planetary companions on orbits with a wide range of semi-major axes (and orbital periods). Even rocky planets (super-Earth and Earth-like planets) associated with nearby very-low-mass stars could be detected with the SKA-VLBI.}
    \label{fig:ska_astrometry}
\end{figure}

The astrometric search for planets consists of measuring the positional displacement (or reflex motion) of the star around the center of mass of the planetary system due to the gravitational attraction of a companion. 
This technique allows for the discovery and characterization of compact binary systems, as well as the detection of substellar companions. 
It also allows for the discovery and characterization of extrasolar planets, provided that the astrometric precision is much better than the amplitude of the reflex motion. For instance, a 5 M$_{J}$ planet on a 3 yr orbit around a Sun-like star at 10 pc will produce a reflex motion of 1 mas. Thus, such a planet can be detected with an astrometric precision better than 100 $\mu$as. Several radio astrometric planet searches have been conducted toward young and low-mass stars, which have resulted in the detection of a few planetary companions \citep[e.g.][]{Curiel2019, Curiel2020, Curiel2022}. 
Figure \ref{fig:ska_astrometry} shows the expected astrometric signal produced by planetary companions with orbital periods of up to 10 yr and semi-major axes up to 2 au, associated with stars of different masses and at several distances.
This figure shows that Jovian-kind planets can be found associated with stars of different masses and at a wide range of distances. For the case of nearby ($<$30 pc) low-mass stars, sub-Neptune, and even super-Earth, planets can be detected.

Radio astrometric observations of young and low-mass stars have been crucial, primarily because they have enabled the determination of their precise trigonometric distances, which are important for determining their luminosity, mass, and ages. In addition, this kind of study provides the dynamical mass of young and low-mass binary systems. These properties are central to understanding the physics of these objects \citep[e.g.][]{Dahn2002, Andrei2011, Dupuy2012, Dupuy2013, Smart2013, Sahlmann2014, Curiel2019, Curiel2024}.

Astrometry uniquely constrains the inclination and position angle of the line of nodes of the host star’s reflex orbit, allowing for the estimate of true companion masses and mutual-inclination measurements when combined with RV/transit data.
In particular, radio interferometric astrometry, with milliarcsecond angular resolution and very high astrometric precision (better than 10 $\mu$as) can be used to search for planetary companions associated with single stars, as well as within binary systems.

The search for exoplanets associated with young, active stars with strong magnetic fields is difficult for most of the current techniques because the stellar activity produces cyclical changes in the host star's photosphere and chromosphere that mask or mimic planetary signals. However, these same stellar characteristics are an advantage for the radio astrometric search for exoplanets because when the magnetic activity of the star is stronger, the radio emission of the star is also stronger. Thus, a search for exoplanets using VLBI radio observations will provide the detection of an exoplanet population that cannot be reached by other techniques. In addition, VLBI monitoring has the potential to demonstrate conclusively the power of the astrometric method to detect exoplanets around sub-stellar hosts. This exoplanet population is largely unknown, and the success of this methodology would open up the brown dwarf regime for planet hunting. Recent results have demonstrated the ability for VLBI astrometry to make a unique contribution to the field of exoplanet research. The detection of planetary companions around very low-mass stars and brown dwarfs will challenge present formation and evolutionary models. Furthermore, the radio astrometric search, and detection, of exoplanets around very low-mass stars (including brown dwarfs) has the potential to provide crucial information for determining the physical mechanisms that rule the formation and evolution of planets formed in the disks around these types of star.

\section{Ultra-precise radio astrometric search for exoplanets with SKA-VLBI} 
The detection of active young and low-mass stars depends on the radio flux density of the star and the sensitivity of the telescope. Since the radio emission is non-thermal in nature and the sources are expected to be variable, deep centimeter observations (obtaining a sensitivity between 1 and 10 $\mu$Jy/beam, in some cases even below the $\mu$Jy/beam limit) are required to guaranty the detection of the stars with high signal-to-noise ratio (S/N). 
However, astrometric precision can depend on systematic errors from the observational and analysis strategies, the angular resolution of the observations, the S/N at which the star is detected, and the astrometric residuals. 
The higher the precision of the astrometric observations, the easier the detection of exoplanets can be. Current VLBI astrometry precision using standard methods is on the order of 100 or 50 $\mu$as at 1.4 or 8.4 GHz, respectively, with angular resolutions of the order of a few milli-arcsec, but SKA-VLBI observations using MultiView will have the potential to reach astrometric accuracies of about 1-10 $\mu$as with a similar angular resolution (see Section \ref{sect:Precision_SKAVLBI}).
Thus, SKA-VLBI observations will allow for the detection of astrometric signals between 10 and 50 times weaker than present VLBI observations, respectively.

The expected astrometric signal is small, from a small fraction of mas ($\gtrsim$10 $\mu$as) to several mas, depending on the mass of the star and its distance, as well as the mass of the planet and its orbital period (or semi-major axis). Figure \ref{fig:ska_astrometry} illustrates the expected capabilities of SKA-VLBI to detect the astrometric signal for some combinations of stars and distances. 
A minimum astrometric signal of 50 $\mu$as ($\sim$5-$\sigma$) was used in all panels of this Figure. However, SKA-VLBI observations are expected to achieve higher astrometric precision. 
This figure shows that for a given star at a given distance, Jovian planets (e.g., Jupiter-like planets) with a wide variety of orbital periods (and semi-major axes) can be found. On the other hand, rocky planets (e.g., super-Earths) can also be found if they orbit around nearby very-low-mass stars (including brown dwarfs) with relatively large orbital periods. High-angular resolution observations (of the order of a few mas) with  SKA-VLBI have the potential to indirectly detect a wide variety of planetary companions (from Earth-like to super-Jupiter), with a wide variety of orbits (semi-major axis from a fraction of au to tens of au, and orbital periods from about several months to several years).

\section{Binary Systems with Planetary Companions}
Only a very small fraction of the detected extrasolar planets (less than 4\%) are known to be associated with stars in binary systems. This is probably, at least in part, due to strong observational biases. For example, the radial velocity (RV) technique is in general limited to binary systems with separations greater than 2$''$, and the transit technique is limited to high inclination angles of the orbital planes of the planets and the binary systems \citep[e.g.,][and references therein]{Marzari2019}. In addition, the presence of a stellar companion has adverse effects on planet formation. It is possible that the stellar companion strongly influences both the formation of planets and their subsequent dynamical evolution.
The formation of planets in binary systems is not well understood. Current planetary formation models take into account only the formation of planets around single stars. In addition, dynamical interactions between stars and/or planets \citep[][]{Nagasawa2008} help explain small and large differences in the inclination angle of close-in giant planets.  The chaotic star formation environment during the accretion phase \citep[][]{Bate2010} and perturbations from a stellar companion \citep[][]{Thies2011,Batygin2013,Lai2014} also help to explain a large misalignment between stellar spin and planetary orbit. Only a small fraction (about 15\%) of the planets associated with binaries have been found associated with binary systems with separations $\lesssim$40 au \citep[e.g.,][]{Marzari2019,Fontanive2021}, and according to the Catalog of Exoplanets in Binary Systems \citep[][]{Schwarz2016} only eleven planets are associated with M-dwarf binary systems. 

In such cases, the orientation of the three-dimensional (3D) orbital plane of both the binary system and the planetary companions is very difficult to establish. In most cases, not all angles of the orbital plane of the binary system and their planetary companions are known.
However, astrometry is capable of directly giving the three angles (longitude of the periastron $\omega$, position angle of the line of nodes $\Omega$, and  inclination angle $i$) of the orbital planes. In particular, radio interferometric astrometry, with milliarcsecond angular resolution and very high astrometric precision (usually better than 60 $\mu$as) can be used to search for planetary companions associated with binary systems with wide orbits, as well as close binary systems with separations $\lesssim$40 au.
The expected high astrometric precision and sensitivity of SKA-VLBI will open up the possibility to search for planetary companions associated with binary systems with wide and compact orbits.

\begin{figure}[ht]
    \centering
	\includegraphics[width=0.45\columnwidth]{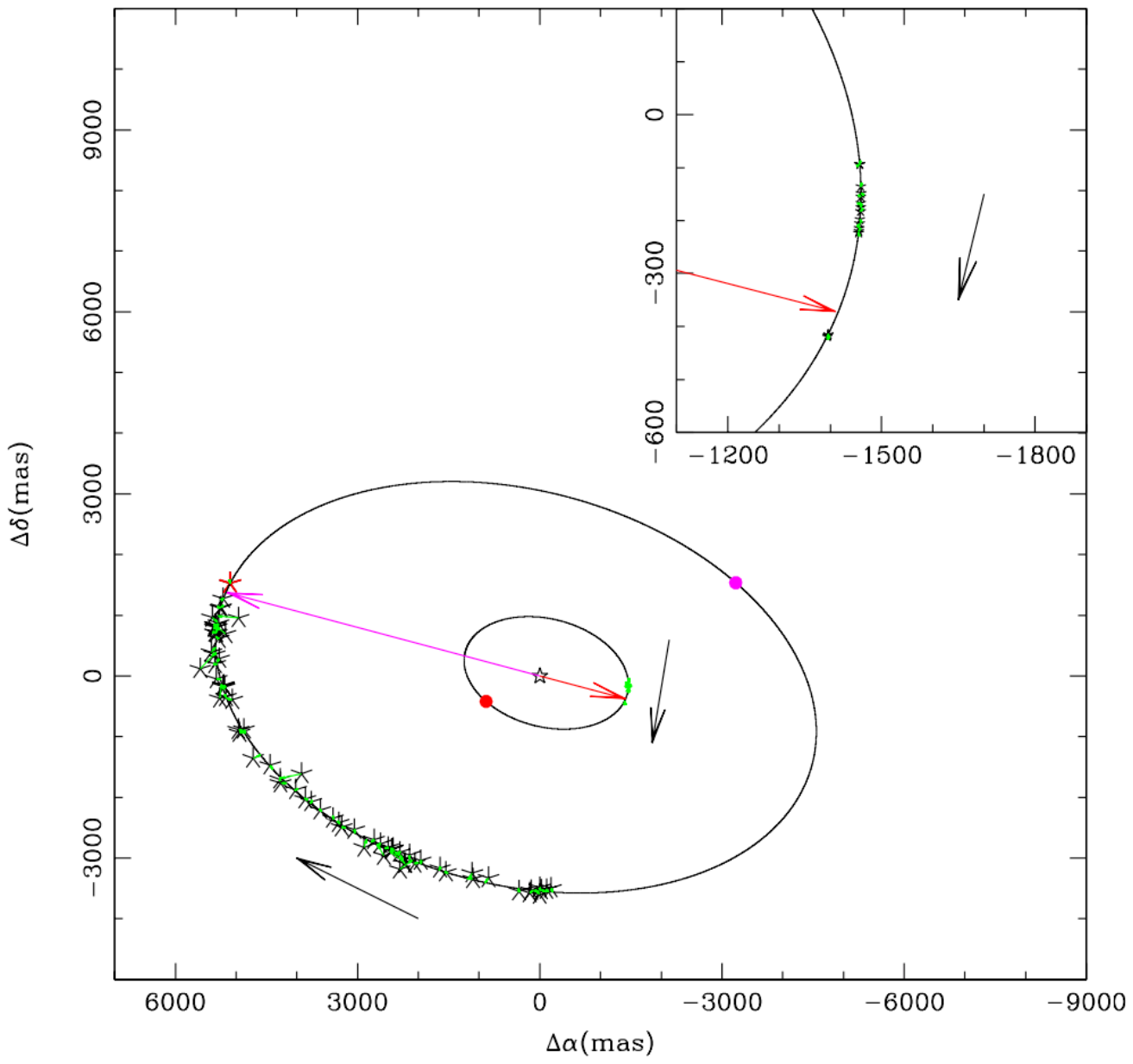}
	\includegraphics[width=0.45\columnwidth]{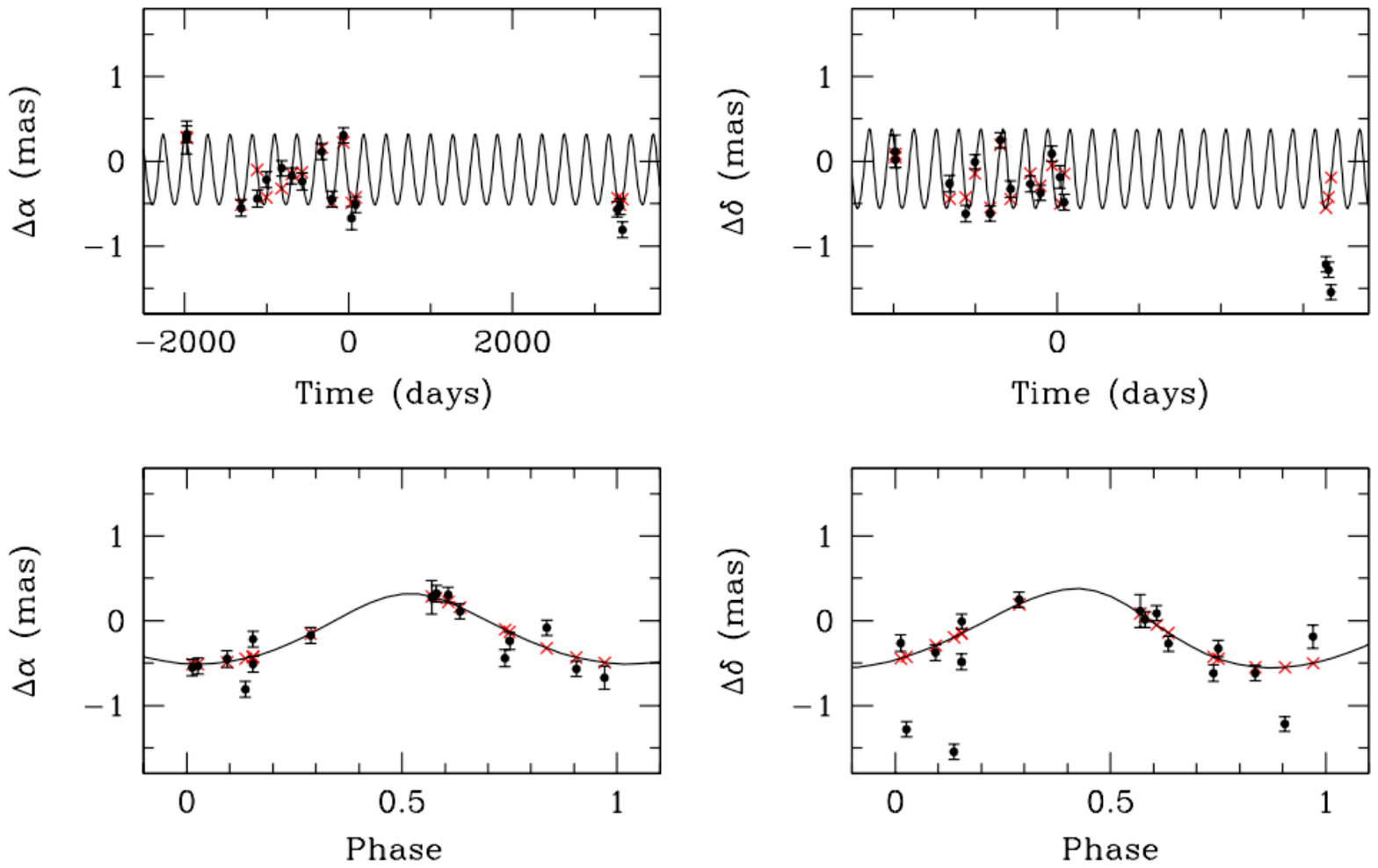}
     \caption{Best-fitted solution using the full combined astrometric fit of the M-dwarf binary system GJ~896AB. 
     {\bf (Left panel:)} Orbital motion of the stellar companion GJ 896B around the main star GJ~896A (big ellipse) and the orbital motion of the main star GJ896A around the center of mass (inner ellipse). The black pointed stars show the observed optical/infrared position, and the two red pointed stars show the observed VLBA positions. The green lines connect the observed positions of the binary system with the expected position along the orbit at each observed epoch.
     {\bf (Right panel:)}  Best-fitted solution including the astrometric signal due to the planetary companion GJ~896Ab, obtained from the full combined astrometric fit. The upper panels show the astrometric signal of the main star GJ~896A due to the planetary companion as a function of time, with the parallax and proper motion subtracted, and taking into account the orbital motion of the stellar companion GJ~896B around the main star GJ 896A. The lower panels also show the astrometric signal of the main star GJ896A due to the planetary companion, but in this case the signal is folded in phase.
    Figure taken from \citet[][]{Curiel2022}.}   
    \label{fig:relative_binary}
\end{figure}

\subsection{Wide Binary and multiple Systems}
\begin{figure}[ht]
    \centering
	\includegraphics[width=0.45\columnwidth]{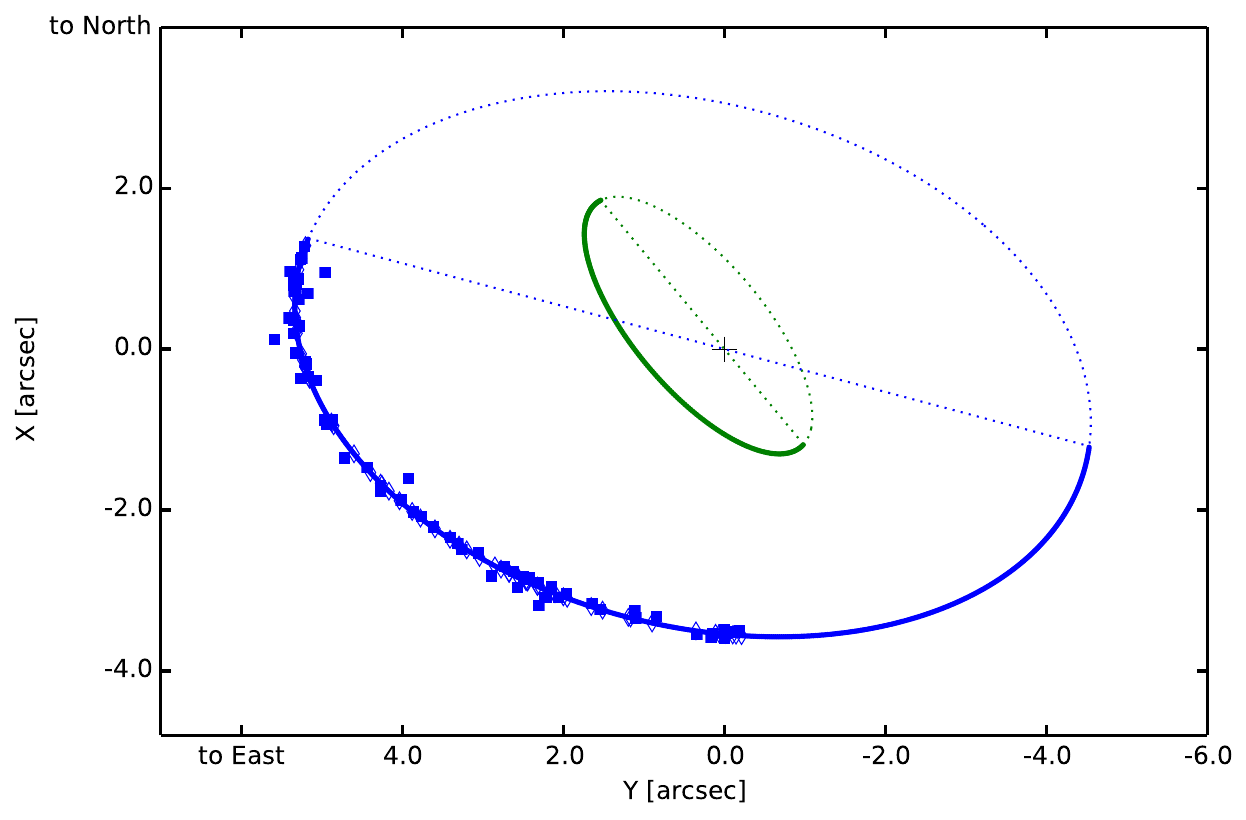}
	\includegraphics[width=0.45\columnwidth]{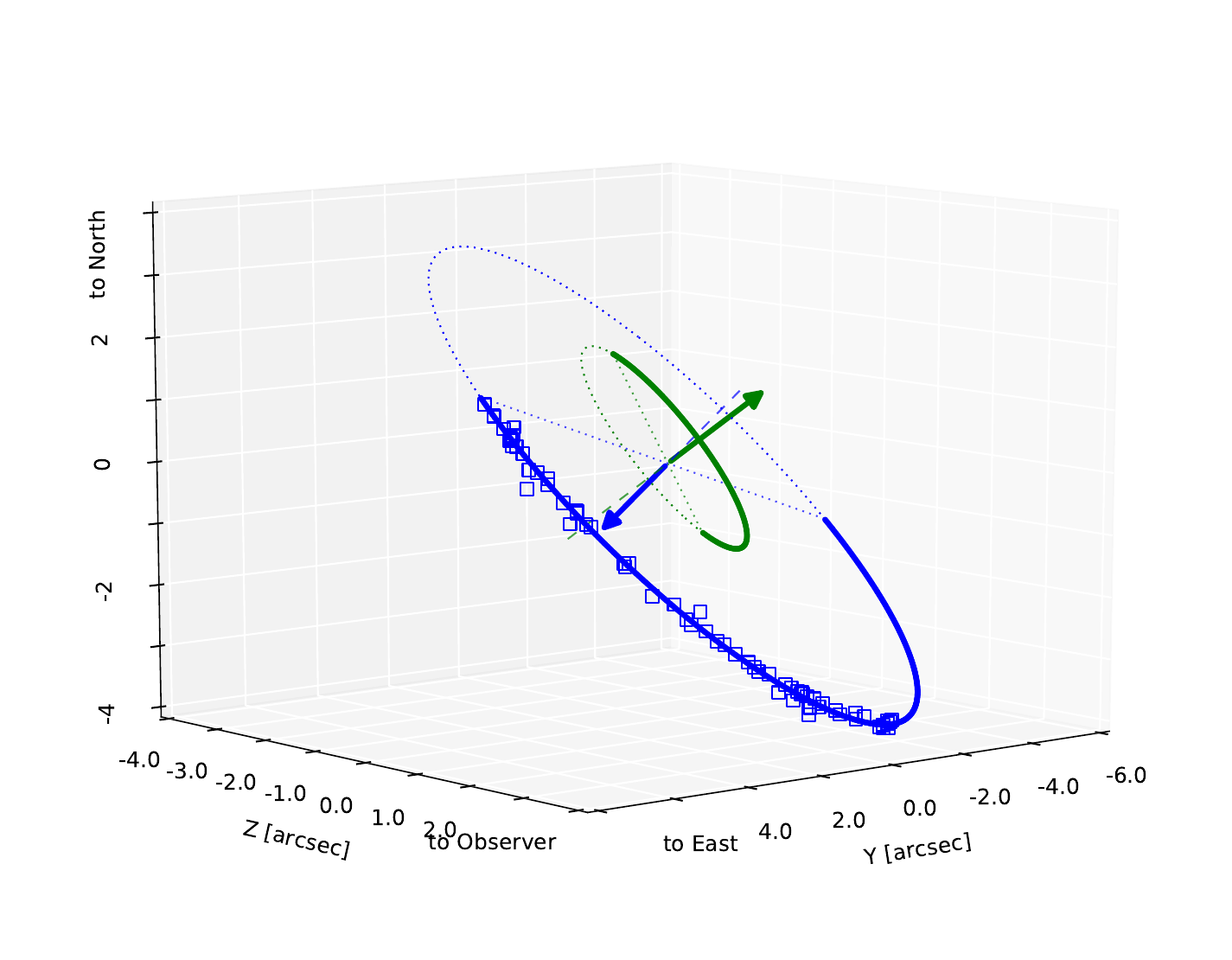}
     \caption{3D orbital architecture of the M-dwarf  binary system GJ~896AB and the planetary companion GJ~896Ab. 
     {\bf (Left panel:)} 2D plot of the fitted orbits of the binary system GJ 896AB (blue) and the planetary companion (green) projected on the plane of the sky. The orbit of the planet has been scaled by a factor of 20 to make the comparison easier. The blue squares show the observed relative position of GJ896B around the main star GJ~896A. The thick lines indicate the side of the orbit that is closer to us, and the dotted lines indicate the side of the orbit that is on the other side of the plane of the sky. The straight dotted lines show the line of nodes of both orbits. 
     {\bf (Right panel:)}  3D plot of the fitted orbits of the binary system GJ 896AB (blue) and the planetary companion GJ 896Ab (green). The orbit of the planet has been scaled by a factor of 20 to make easier the comparison. The arrows indicate the direction of the rotation axis of both orbits.
    Figure taken from \citet[][]{Curiel2022}.}   
    \label{fig:wide_binary}
\end{figure}

Characterizing the full 3D orbital architecture of binary systems containing a planetary companion can aid to investigate the importance of the star-star and star-planet mutual interaction. Combining relative and absolute astrometric data of the binary system, it is possible to obtain the 3D orbital architecture of the system, including the planetary companion. As the full combined astrometric fit (relative plus absolute astrometry) provides the inclination angle and the position angle of the line of nodes of the orbital planes of both the planetary companion and the binary system, the mutual inclination angle of the system can be directly measured.
For example, Figures \ref{fig:relative_binary} and \ref{fig:wide_binary} show the results of the combined absolute and relative astrometric fit of the M-dwarf binary system GJ~896AB with an orbital period and a separation of about 229 yr and 31.6 au \citep[][]{Curiel2022}. The combined astrometric fit also revealed a Jovian planetary companion (m$_\mathrm{p}$ = 2.26$\pm$0.57 M$_\mathrm{J}$) associated with the main star GJ~896A with an orbital period and semi-major axis of 284 days and 2.26 au. In addition, a large mutual angle $\Phi$ = 148$^{\circ}$ was found between the orbits of the binary system and the planet around its host star, which is consistent with the orbit of the planet being retrograde.

SKA-VLBI includes the possibility that all members of a binary or multiple system could be detected in a single observation using multiple TABs for SKA and multiple phase centers for the other antennas in the array (see Section \ref{sect:Precision_SKAVLBI}). This opens up the possibility of searching for planetary companions in all members of binary and multiple systems.

\subsection{Compact Binary Systems}

\begin{figure}[ht]
    \centering
	\includegraphics[width=0.45\columnwidth]{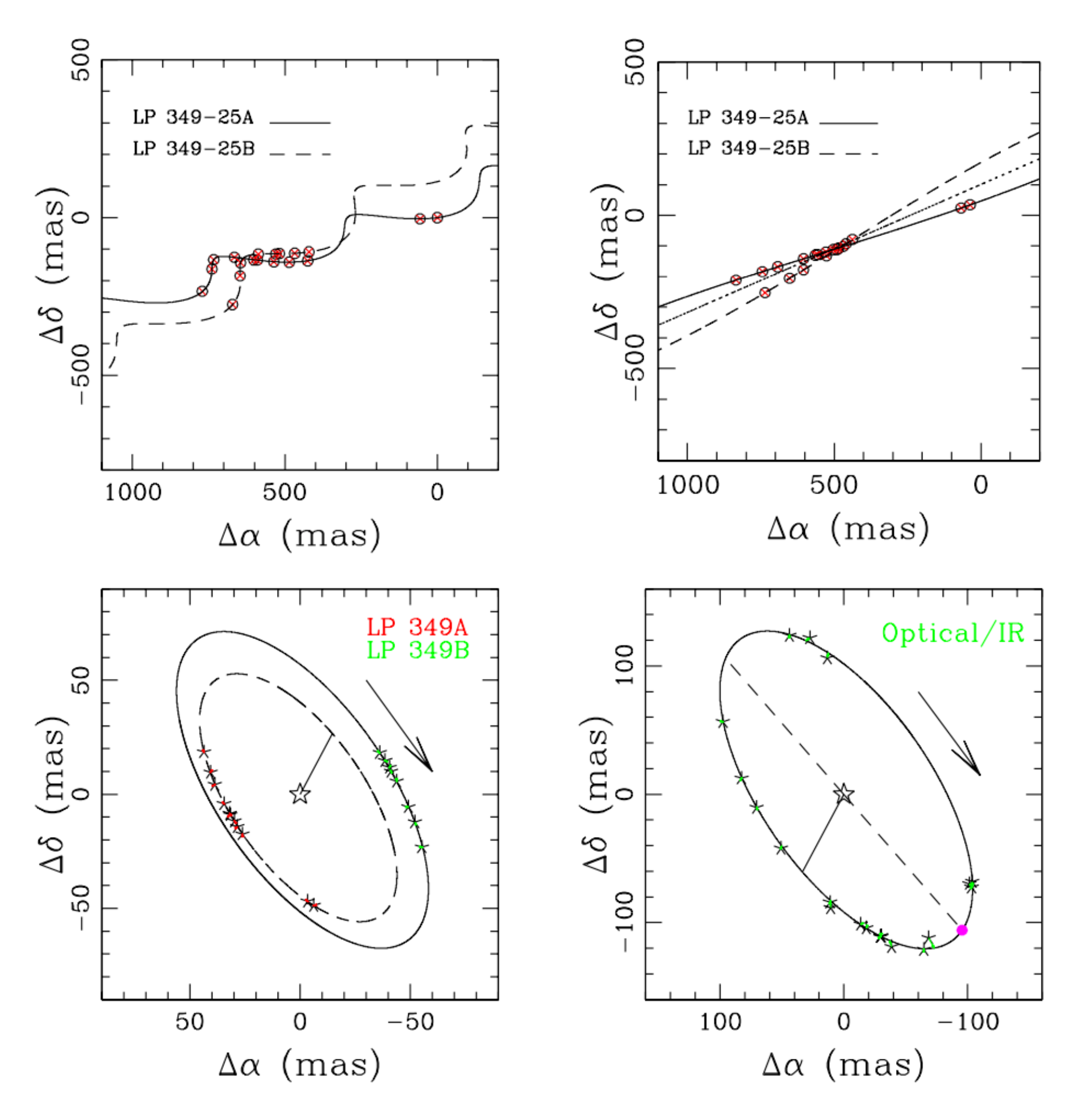}
     \caption{{\bf (Upper-Left panel:)} Absolute astrometric fits of the astrometric positions of UCD binary system LP~349$-$25AB, obtained with the VLBA. The fits includes only proper motions, parallax and the orbital motion of both stars around the barycenter of the binary system. {\bf (Upper-Right panel:)} Same as upper-left panel but removing the contribution of the parallax. The dotted line shows the trajectory (from NW to SE) of the barycenter of the binary system.
    {\bf (Lower-Left panel:)} Orbital motion of both UCD stars around the barycenter of the binary system. The VLBA observations cover approximately 20$\%$ of the orbit. 
    {\bf (Lower-Right panel:)} Relative orbital motion of the secondary star LP~349$-$25B around the primary star LP~349$-$25A. The optical/infrared observed epochs are shown in green. 
    Figure taken from \citet[][]{Curiel2024}.}   
    \label{fig:vlba_binary}
\end{figure}

Compact binaries with separation $<$40 mas cannot be resolved with ground-based optical/infrared telescopes, not even with space telescopes. The best optical/infrared telescopes, such as VLTI/GRAVITY, can resolve binaries with separation larger than $\gtrsim$40 mas,
but only when the projected separation of the binary is close to its maximum. The expected milli-arcsecond resolution with SKA-VLBI will allow to resolve compact binaries. Multi-epoch observations of compact binaries will allow to obtain high-precision estimates of the distance and the masses of their components with expected errors of the order of $\sim$1-3$\%$, or even better  \citep[e.g.][]{Dupuy2016,Azulay2017,Curiel2024,Curiel2026}, which cannot be achieved even with Gaia because this kind of binaries are unresolved and their photo-center is located somewhere between the two stars, depending on the mass ratio and flux ratio of the components. 
For example, Figure \ref{fig:vlba_binary} shows the results of combined absolute and relative astrometric observations of the compact UCD binary system LP~349-25AB \citep[][]{Curiel2024}. The combined fit of this system provided not only the proper motions, parallax and orbital motion of this binary, but also the dynamical mass of the binary and the individual components, which allowed to test stellar evolutionary models. The astrometric solution shows that one of the components is an UCD (85.71$\pm$0.64 $M_\mathrm{J}$), while the estimated  mass of the second component is consistent with being a BD (67.11$\pm$0.51 $M_\mathrm{J}$).

Multi-epoch SKA-VLBI observations of compact binary systems will open the possibility to: 
\begin{itemize}
   \item Obtain high-precision astrometric distances of the binaries, which are necessary to improve the physical properties of the system and the components.
   \item Obtain high-precision dynamical masses of the system and the individual components. This is particularly important in the case of very low-mass binary systems, where the mass of at least one of the components could be close, or even below, the hydrogen-burning limit and thus be BDs \citep[e.g.][]{Curiel2024}. These kind of binaries are poorly studied, and radio astrometry is the best way to obtain their masses.
   \item Calibrate Formation and Evolutionary Track Models. Since not many compact low-mass binary systems have dynamical mass estimates, these models are still poorly constrained \citep[e.g.][]{DupuyLiu2017}. Extending the growing population of UCD binary systems will provide new systems for refining evolutionary theories at the lowest stellar masses into the substellar regime.
   \item Search for close-in planetary companions associated with one or both stars in compact binary systems \citep[e.g.,][]{Curiel2022}. 
\end{itemize}

\section{Direct detection of exoplanets (planetary companions and JuMBOs)}

The possible detection of coherent radio emissions associated with planetary systems, or directly associated with planetary companions, has motivated searches from MHz to GHz frequencies because of their potential to probe the unknown magnetic properties of exoplanets  \citep[e.g.][]{Zarka2007,Hallinan2013,Pineda2023}. Direct detection of a planet at radio frequencies is expected to be very difficult from the ground because the known mechanisms would result in emission at very low frequency (below the low frequency end of the radio window set by the Earth’s ionosphere) or very weak \citep[e.g.][]{Zarka2007,Katarzynski2016}. In the cases where the magnetic field on the surface of the star is strong (B$_{*}$ $>$ 1 kG), the star-planet interaction could produce detectable radio emission at centimeter wavelengths, in a way equivalent to the Jupiter-Io interaction, with the star playing the role of Jupiter and the planet that of Io  \citep[e.g.][]{Zarka2007}. 

Recent searches with LOFAR have resulted in the detection of coherent low-frequency radio emission ($\lesssim$200 MHz) from several early M-dwarfs \citep[][]{Vedantham2020,Callingham2021,Zhang2025}. However, these stars need confirmation that a planetary companion indeed drives the radio emission. Some of these stars have been observed with the VLA but were not detected \citep[e.g.][]{Ortiz-Ceballos2024}. Recent radial velocity (RV) studies have revealed that one of the stars detected with LOFAR (GJ 1151) has two low-mass planetary companions \citep[e.g.][]{Blanco2023}. However, the characteristics of these planets cannot explain the origin of the radio emission of this slowly rotating M4.5 star. GJ 1151 was observed with the VLA but was not detected \citep[][]{Ortiz-Ceballos2024}.
The unprecedented sensitivity of SKA-low will help to better characterize the MHz radio emission from evolved planetary systems. Furthermore, SKA-low surveys will provide many more candidates for star-planet interactions. On the other hand, the astrometric precision, angular resolution, and sensitivity of SKA-VLBI observations of planetary systems will help to identify the origin of the radio emission.

VLA and ATCA centimeter observations of planetary systems have provided a few candidates to study star-planet magnetic interaction \citep[][]{Perez2021,Pineda2023,Ortiz-Ceballos2024}. YZ Cet, with three small planets in a compact configuration, was detected with the VLA, suggesting that the observed coherent radio bursts could be associated with the star-planet interaction \citep[][]{Pineda2023}. Polarized radio emission from Proxima Centauri, detected with ATCA, exhibits a possible orbital periodicity with Prox Cen b \citep[][]{Perez2021}, but the planet’s 11-day period places it at an orbital distance unlikely to have sub-Alfv\'enic interaction \citep[][]{Kavanagh2021}. Moreover, the possibility of coherent radio bursts entirely of stellar origin remains a significant possibility as magnetically active M-dwarf stars frequently exhibit polarized radio emissions \citep[][]{Lynch2017,Villadsen2019}, and the slowly rotating M-dwarf Prox Cent exhibits stellar flare-associated coherent radio bursts \citep[][]{Zic2020}. Radio studies of young and low-mass stars have not yet provided the direct detection of planetary companions. Recent searches of radio emission from low-mass stellar systems hosting planets with the VLA have provided the detection of only one M-dwarf (GJ 3323) at centimeter wavelengths  \citep[][]{Ortiz-Ceballos2024}. This is an excellent candidate for investigating the origin of radio emission in planetary systems. However, this star has not been observed with the VLBA, while VLBI observations are the best way to find the origin of the radio emission (the star, the planet or both). The closest result so far is the direct detection of a low-mass brown dwarf (a 35 Jupiter masses companion) orbiting the young WLTT star DoAr21  \citep[][]{Curiel2019}. The expected higher astrometry precision and sensitivity with SKA-VLBI will open the possibility not only of detecting many more candidates but also of finding the origin of the radio emission, which could be associated with the star, the planetary companion or both. SKA-VLBI has the potential of detecting radio emission from planets in planetary systems, which is crucial for understanding the magnetic fields in exoplanets. Furthermore, in the cases where the star and the planet were detected, multi-epoch SKA-VLBI observations would provide the dynamical mass of both the star and the planet. 
Detailed discussions about star-planet interaction can be found in other chapters in the Advance Astrophysics with the Square Kilometre Array II, such as magnetic star-planet interaction {\color{black}(Radio emission from star-planet interactions by \citealp[][]{Vedantham01.2026.SKA}}), and direct exoplanet detection {\color{black}(Discovering and characterising exoplanets and ultracool dwarfs with the Square Kilometre Array by \citealp[][]{Kavanagh01.2026.SKA})}.

The search for interstellar $''$rogue planets$''$ (single planets not gravitationally bound to a star), also known as free-floating planets, is an alternative method of finding planetary bodies that are not associated with stars. An increasing number of free-floating planets have been detected in the optical and near-infrared in the past few decades   \citep[e.g.,][]{Lucas2000,Zapatero2000,McCaughrean2023}. Very few candidates for free-floating planetary-mass objects have been discovered using radio observations \citep[e.g.,][]{Kao2018}. Recently, a Jupiter-mass binary-object (JuMBO), discovered by observations with the James Webb Space Telescope \citep[][]{McCaughrean2023}, has been detected at radio wavelengths with the VLA. However, due to the low angular resolution of the VLA observations and the lack of detection with the HSA \citep[][]{Rodriguez2024,Rodriguez2025}, the radio emission of this planetary binary system was not spatially resolved, and it is not clear if the emission is associated with one or both planetary-mass bodies. In addition, the nature of the radio emission of the JuMBOs is not clear. 
SKA-Mid will probably be able to detect many of the JuMBOs. Higher angular resolution observations will be needed to obtain the orbital structure of JuMBO systems, and
the expected astrometry precision, sensitivity, and field of view with SKA-VLBI will be key to obtain their dynamical masses and to find the origin of the radio emission of the JuMBOs.

\section{Synergies with other telescopes and techniques: Gaia, ngVLA, RV, Transit}
\subsection{Gaia, ngVLA}
Gaia’s astrometric observations have the potential to detect many (probably thousands) exoplanets and brown dwarfs associated with solar-type and low-mass stars \citep[e.g.,][]{Casertano2008,Perryman2014,Sozzetti2014},
preliminary results have been published \citep[][]{Holl2023,Gaia2023,Stefansson2025}.
Gaia's DR4 release, due in late 2026, is expected to include the first publication of temporal astrometry (5.5 years of observations) and an initial catalog of exoplanets. The full list of exoplanets detected with Gaia will be published after 2030, when the final Gaia catalog DR5 (10 years of observations) is expected to be published. However, Gaia is not sensitive to low-mass magnetically active stars, which are more suitable to study with radio observations. In addition, radio VLBI observations are better suited to resolve and study compact binary systems. Thus, the search for exoplanets with Gaia and SKA-VLBI is complementary. The new technique of radio astrometry is gaining importance and will be important even after the final publications of Gaia's temporal data.
Furthermore, in the case of sources detected by both SKA-VLBI and Gaia, combining both datasets will increase the temporal astrometric monitoring of radio sources, helping to search for planetary companions in very wide orbits, with orbital periods of more than 10 years.

SKA-VLBI and ngVLA  are ideal for searching for exoplanets using radio astrometry. Both interferometers will cover part of the same frequency range, suitable  for this technique. ngVLA will observe sources in the northern hemisphere, while SKA-VLBI will observe mainly sources in the southern hemisphere, therefore, they will complement each other and together cover the whole sky. This will considerably increase the number of exoplanets that could be detected using radio astrometry.

\subsection{Other techniques such as RV and Transit}
The techniques of Radial Velocity and Transit have been very successful in finding thousands of planetary companions with a wide range of orbital structures. However, these techniques are more susceptible to compact orbits, and their effectiveness decreases with the orbital distance of the companion. Astrometry, on the other hand, is more sensitive to companions in wide orbits and its sensitivity increases with the orbital separation of the companion. This makes astrometry a complementary technique capable of detecting Jupiter-like  planetary companions at several astronomical units from the host star, in orbits similar to the Jovian planets in the Solar System. 

Astrometry provides all the Keplerian orbital parameters of the substellar and planetary companions, including the three angles (longitude of the periastron $\omega$, position angle of the line of nodes $\Omega$, and inclination angle $i$) of the orbital planes. However, there is an ambiguity in the position angle of the ascending node ($\Omega$ and $\Omega+180^{\circ}$) that cannot be removed by absolute astrometric observations alone. 
This ambiguity can be removed by combining astrometry and RV techniques \citep[e.g.,][]{Curiel2024}.

\section{Final Remarks}
The main focus of this chapter has been on the detection and study of extrasolar planets using SKA-VLBI radio astrometry. While currently only a few exoplanets have been detected with present radio interferometers, next generation SKA-VLBI interferometer and next generation MultiView methods  will provide an order of magnitude improvement in sensitivity and astrometric performance that will open the possibility of detecting hundreds, even thousands, of exoplanets associated with  different kinds of stars (e.g., young and low-mass stars, including brown dwarfs). The search for exoplanets at radio wavelengths will be complementary to other techniques, including the astrometric search for exoplanets with Gaia.  In addition, the search for exoplanets using SKA-VLBI radio observations will provide the detection of an exoplanet population that is difficult to reach by other techniques.

~~~~~

{\bf Acknowledgements}

S.C. acknowledges financial support from UNAM and Secretaría de Ciencia, Humanidades, Tecnolog{\'\i}a e Innovaci\'on (Secihti), M\'exico. 
S.C. acknowledges financial support from UNAM-PAPIIT IN107324 grant and from Secihti grant number CF-2023-I-232. 

{\it Software used:} python3.10, numpy, matplotlib

\bibliography{chapter} 

\bibliographystyle{abbrvnat-maxbibnames4}

\end{document}